# Density of hybrid plasma generated by microwave and laser radiation in the Ar:$H_2$:$CH_4$ mixture

S.V. Avtaeva[1,2], V.B. Dolomanova[1,2], P.A. Pinaev[1], A.E. Medvedev[1]

[1]*Institute of Laser Physics SB RAS, 630090, Novosibirsk, Lavrentieva Ave., 15B,*

*avtaeva_sv@laser.nsc.ru*

[2]*Novosibirsk State Technical University, 630073, Novosibirsk, K. Marx Ave. 20*

**Abstract.** The atmospheric-pressure hybrid plasma in the Ar:$H_2$:$CH_4$ mixture, maintained by microwave radiation (2.47 GHz) and $CO_2$ laser radiation (10.6 μm) in the chamber of an experimental plasma-chemical reactor designed to study the synthesis of diamond-like coatings was studied. The electron number density was determined from the Stark broadening of the $H_α$ line shape of atomic hydrogen. It was shown that when laser radiation is focused in the region of a microwave plasma bunch, the $H_α$ line shape in the hybrid plasma spectra has broad wings and is described by a Lorentz function with a two-contour approximation. The complex structure of the $H_α$ line profile of the hybrid plasma indicates its spatiotemporal inhomogeneity. The electron number density corresponding to the contour with a smaller half-width exceeds the electron number density in microwave plasma and lies in the range of $(4-8) \cdot 10^{15}$ cm$^{-3}$, and the electron number density measured by the contour with a larger half-width is $\sim (1.5-2) \cdot 10^{17}$ cm$^{-3}$.

Diamond-like polycrystalline films possess a number of useful properties, such as exceptional hardness and wear resistance, low friction coefficient, chemical inertness, biocompatibility, and optical transparency (including UV and IR ranges) [1]. These properties provide a wide range of applications for diamond-like films: in mechanical engineering, transportation, medicine, optics, and the electronics industry. CVD synthesis of diamond-like coatings using microwave plasma has become widespread. The high homogeneity and stability of the characteristics of a nonequilibrium microwave discharge, at gas pressures significantly below atmospheric, allows for the production of high-quality materials [2, 3]. However, the characteristic growth rates of polycrystalline diamond coatings in such microwave plasmas typically do not exceed tens of microns per hour [4]. Limitations in the growth rate of diamond-like films are due to the relatively low density of charged particles in microwave plasmas at low and medium pressures and the correspondingly low rate of production of chemically active plasma components necessary for coating synthesis [4, 5].

A significantly higher density of charged particles, which could ensure a higher growth rate of the synthesized layer, can be achieved in laser plasma [6]. At atmospheric pressures, the characteristic values of the charged particle density and the electron temperature for $CO_2$ laser plasma are $10^{15}$-$10^{17}$ cm$^{-3}$ and 1-2 eV, respectively [7, 8]. However, laser plasma is characterized by significant non-uniformity and a small synthesized surface area [9]. Furthermore, laser plasma typically has a high gas temperature, which, during the longest relaxation phase after the laser pulse is applied, approaches the electron temperature. This may limit the application of laser plasma, since, for example, the substrate temperature during diamond film synthesis should not exceed 1000K [10, 11]. A significant increase in the density of charged particles while maintaining the homogeneity of the non-equilibrium plasma can be expected by introducing pulse-periodic $CO_2$ laser radiation into atmospheric pressure microwave plasma [12].

In this paper, we investigate the number density of electrons in a hybrid plasma at atmospheric pressure (1.2 atm) in a gas mixture of 89Ar:10H$_2$:1CH$_4$, supported by microwave radiation (2.47 GHz) and a $CO_2$ laser radiation (10.6 μm) in the chamber of an experimental plasma-chemical reactor developed for studying the synthesis of diamond-like coatings [13]. Hybrid plasma was generated in the reactor chamber (Fig. 1), which is a quasi-cylindrical microwave resonator modeled on the basis of the $TM_{011}$ mode.

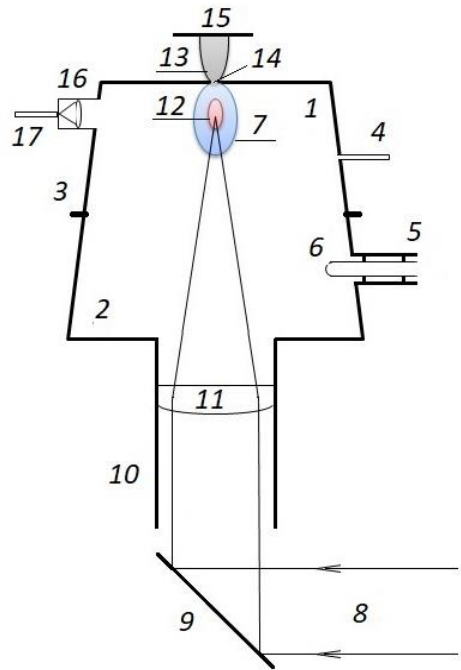

Fig. 1. Schematic diagram of hybrid plasma formation in the chamber of the experimental reactor. *1, 2* - small and large cones of the chamber, *3* - connection area of the chamber cones, *4* - gas inlet, *5, 6* - microwave cable and pin for inputting mw radiation into the chamber, *7* - microwave plasma , *8* - $CO_2$ laser radiation, *9* - copper mirror, *10* - evanescent waveguide, *11* - long-focus lens, *12* - hybrid plasma, *13* - jet of hybrid plasma products, *14* - nozzle, *15* - substrate, *16* - collimator for feeding of plasma radiation into optical fiber *17*.

The small cones *1* and large cones *2* of the reactor chamber are connected by a thread *3*, allowing the volume to be adjusted for resonance with the microwave generator. A gas mixture (argon, hydrogen, and methane) is fed into the resonator chamber through inlet *4*, creating excess pressure. The composition of the gas mixture is determined by a four-channel gas mixer, which controls the gas flow rate in each channel. When microwave radiation is fed into the resonator via input pin *6*, gas breakdown occurs in the region of the main maximum of the EM standing wave, forming microwave plasma *7*. The radiation from the $CO_2$ laser *8* is directed by a copper mirror *9* through an evanescent waveguide *10* into the reactor chamber, where it is focused by a lens *11* (ZnSe, $d = 50$ mm, $F = 250$ mm) into a microwave plasma bunch, in which a hybrid plasma *12* is formed. A jet of plasma products *13* exits the reactor through a nozzle *14* with a cross-section of 1 mm in diameter and is directed toward a substrate *15*, where the deposition takes place. A $CO_2$ laser system with an average power of up to 2 kW, operating in a pulse-periodic mode, produces pulses of up to 100 kW. The laser pulse repetition rate can be set between 1 and 150 kHz, and the pulse duration is approximately 1 μs. The reactor can operate both in the microwave discharge mode, when only microwave radiation is supplied to the chamber, and in the hybrid discharge mode, when microwave and laser power are simultaneously supplied. Plasma radiation is collected by a collimator *16* with an input aperture of 15 mm diameter, enters the light guide *17* and then into the HDSA spectrum analyzer (Ångstrom, Novosibirsk, Russia & HighFinesse, Tübingen, Germany) with a resolution of 0.2-0.3 Å depending on the wavelength (350-1000 nm).

When microwave radiation was applied, microwave plasma was generated in the resonant reactor, which could be detected by the characteristic gas glow in the reactor window. After igniting the microwave discharge, radiation from the $CO_2$ laser operating in a pulsed-periodic mode was injected into a microwave plasma bunch approximately 20 mm in size, centered on the chamber axis, at a depth of ~10 mm from the nozzle. The laser radiation was focused approximately at the center of the microwave plasma bunch; the diameter of the laser radiation waist was 100 μm. When the $CO_2$ laser radiation was injected into the microwave plasma bunch, a bright flash of radiation was visually observed inside the bunch. The frequencies $v$ and the radiation power $\bar{P}$ of the laser pulses used, linked by the relation $\bar{P} = E_{pulse} \cdot v$, are given in Table 1 ($E_{pulse}$ is the pulse energy). The microwave power supplied to the discharge was 880 W at a pulse repetition rate of 10 kHz and a duty cycle of 5.

The microwave discharge spectrum exhibits lines of atomic hydrogen of the Balmer series ($H_\alpha$, $H_\beta$, $H_\gamma$), Swan bands of molecular carbon ($C_2$), and atomic argon lines. In the hybrid plasma spectrum, the lines of atomic hydrogen of the Balmer series and atomic argon were highly intense, while the intensity of the Swan bands varied only slightly. Furthermore, a large number of $Ar^+$ ion lines were observed in the hybrid plasma spectrum in the wavelength range of 380–520 nm (Fig. 2).

Table 1. $CO_2$ laser radiation parameters ($\lambda$=10.6 μm).

| Pulse repetition frequency, kHz | Average power, W | Pulse energy, mJ |
|---|---|---|
| 3 | 140 | 46,7 |
| 15 | 370 | 24,7 |
| 30 | 520 | 17,3 |
| 60 | 670 | 11,2 |

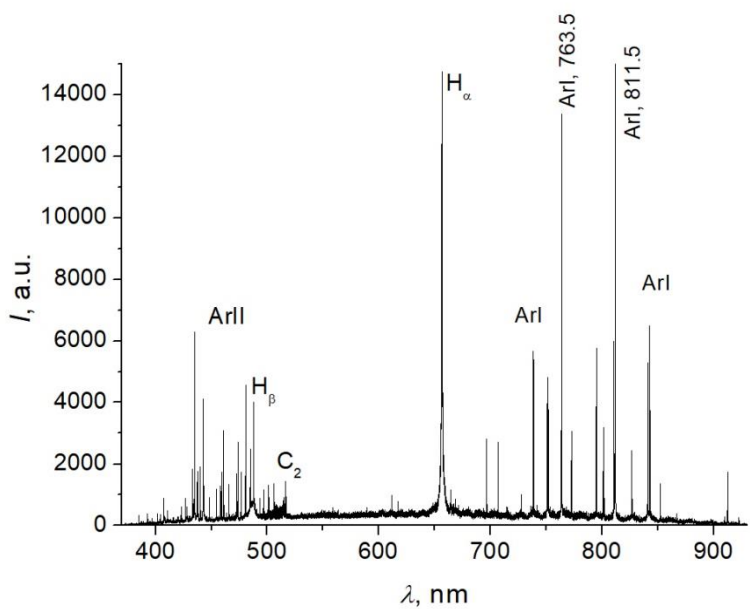

Fig. 2. Typical emission spectrum of the hybrid plasma in the mixture of 89Ar:10H$_2$:1CH$_4$ at a laser pulse repetition rate of 30 kHz.

According to the literature, the electron density in microwave discharge plasma at atmospheric pressure is $10^{13}$-$10^{14}$ cm$^{-3}$, which leads to a noticeable broadening of the atomic hydrogen lines due to the Stark effect, which is used to measure the electron density [4, 5]. The H$_\beta$ line is most convenient for determining the electron density [14]. However, in the emission spectrum of the microwave discharge in the 89Ar:10H$_2$:1CH$_4$ mixture, the H$_\beta$ line intensity was low, and in the case of hybrid plasma, several lines of argon Ar$^+$ ions overlapped the H$_\beta$ line, making it difficult to measure its shape. Therefore, the H$_\alpha$ atomic hydrogen line was used to measure the electron density. The instrumental function of the recording spectral system was measured using a helium-neon laser. Analysis of the instrumental function shape showed that it is well approximated by a Gaussian function with a half-width of $\Delta\lambda_a = 0.02$ nm. The contribution of Doppler broadening to the $H_\alpha$ line shape was estimated from the measured rotational temperature of $C_2$ molecules, which, under the conditions under consideration, is close to the gas temperature. According to estimates, the combined instrumental and Doppler broadening of the recorded $H_\alpha$ line shape in hybrid plasma, $\Delta\lambda_G$, does not exceed 0.03 nm.

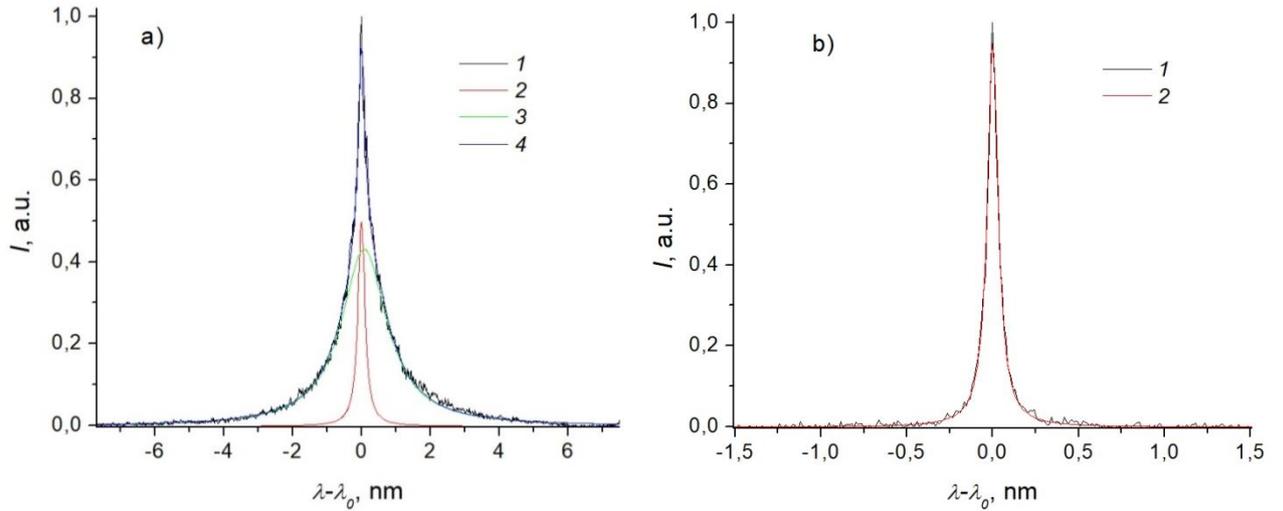

Fig. 3. The shape of the Hα line observed in the radiation of hybrid and microwave discharges in the gas mixture 89Ar:10H$_2$:1CH$_4$; a) hybrid plasma at the CO$_2$ laser pulse repetition rate of 15 kHz, *1* - measured shape, *2, 3* - narrow and wide approximation contours, *4* - two-contour Lorentzian approximation; b) microwave plasma, *1* - measured shape, *2* - approximation by the Voigt function.

Figure 3 shows the shape of the $H_\alpha$ line observed in the plasma of the hybrid and microwave discharge. The profile of the $H_\alpha$ line emitted by the hybrid plasma (Fig. 3a) is much wider than its combined instrumental and Doppler broadening; therefore, when estimating the Stark broadening of the $H_\alpha$ line, the instrumental and Doppler broadening can be neglected. The profiles of the $H_\alpha$ line emitted by the hybrid plasma are well approximated by two Lorentz functions (two-contour approximation) and cannot be approximated by either a single Lorentz function or the Voigt function. A typical form of the two-contour Lorentzian approximation of the $H_\alpha$ line profiles is shown in Fig. 3a. The shape of the $H_\alpha$ line emitted by microwave plasma has a smaller width and is well approximated by the Voigt function (Fig. 3b). The Stark contribution to the half-width of the H$_\alpha$ line shape was extracted from the experimental shape using the tabular relations $\Delta\lambda_{1/2}^L/\Delta\lambda_{1/2}$ and $\Delta\lambda_{1/2}^G/\Delta\lambda_{1/2}$ (Table 6.5 in [15]), where $\Delta\lambda_{1/2}$ is the observed half-width of the experimental shape, $\Delta\lambda_{1/2}^L$ is the Lorentz half-width due to the Stark effect, and $\Delta\lambda_{1/2}^G$ is the half-width of the Gaussian shape due to the instrumental and Doppler broadening.

It is known that, unlike microwave plasma, which is relatively uniformly distributed in the region of the standing wave field antinode with a characteristic size on the order of a centimeter, laser plasma is highly inhomogeneous [12]. Gas breakdown by pulse-periodic laser radiation initially occurs in a small region with a characteristic size of hundreds of microns, determined by the focusing of the laser radiation. Yu. P. Raizer identified two possible mechanisms for describing the expansion of laser plasma after breakdown [12]. The first mechanism is due to the heating of the gas by laser radiation and its rapid expansion, accompanied by the propagation of a shock wave and ionization. The second mechanism is associated with the emission of radiation by the heated gas, which heats and ionizes the surrounding gas, creating a "radiation wave."

When any of these mechanisms is realized, the plasma region rapidly expands over time. The electron density is highest in the region of the initial gas breakdown and decreases with distance from it. Thus, the laser plasma density is characterized by strong temporal and spatial nonuniformity, which can be reflected on the shape of spectral line caused by Stark broadening under the action of rapidly changing electric fields associated with the spatiotemporal distribution of charged plasma particles.

In our experiments, laser radiation is absorbed not by a neutral gas, but by the microwave discharge plasma. The presence of initial gas ionization leads to effective absorption of laser radiation by microwave plasma electrons in the beam waist region, resulting in the rapid development of an electron avalanche. When focused laser pulses are applied to the microwave plasma bunch, a rapidly expanding burst of plasma radiation is visually observed.

We note that the recording time of spectral line shapes (from a few seconds to 30 s) with our equipment significantly exceeds the characteristic times of the processes occurring in laser plasma.

Therefore, the experimentally observed $H_\alpha$ line shapes, described by a two-contour approximation using the Lorentz function (Fig. 3b), reflect the averaged characteristics of the spatiotemporal nonuniform distribution of electron density within the hybrid plasma volume.

Figure 4 shows the half-width of the narrow and wide contours of the two-contour Lorentz approximation of the experimentally observed $H_\alpha$ line shape and the electron densities measured from the half-width of each of these two contours. The restoration of the electron density from the half-width of each contour of the two-contour Lorentzian approximation of the $H_\alpha$ line was performed using the diagrams obtained in [16]. The electron density calculated from the wide contour of the two-contour Lorentzian approximation is on the order of $(1.5-2) \cdot 10^{17}$ cm$^{-3}$ and apparently reflects the high electron densities in the region of hybrid plasma formation, at the initial moments of time and after the laser pulse.

The electron density calculated from the central narrow contour lies within the range of $(4-8) \cdot 10^{15}$ cm$^{-3}$ and probably corresponds to the peripheral region of the hybrid plasma at later times of the evolution of the hybrid discharge pulse, where the electron density approaches their density in the undisturbed microwave plasma ($\sim 1.5 \cdot 10^{15}$ cm$^{-3}$) as it moves away from the center of the detonation wave.

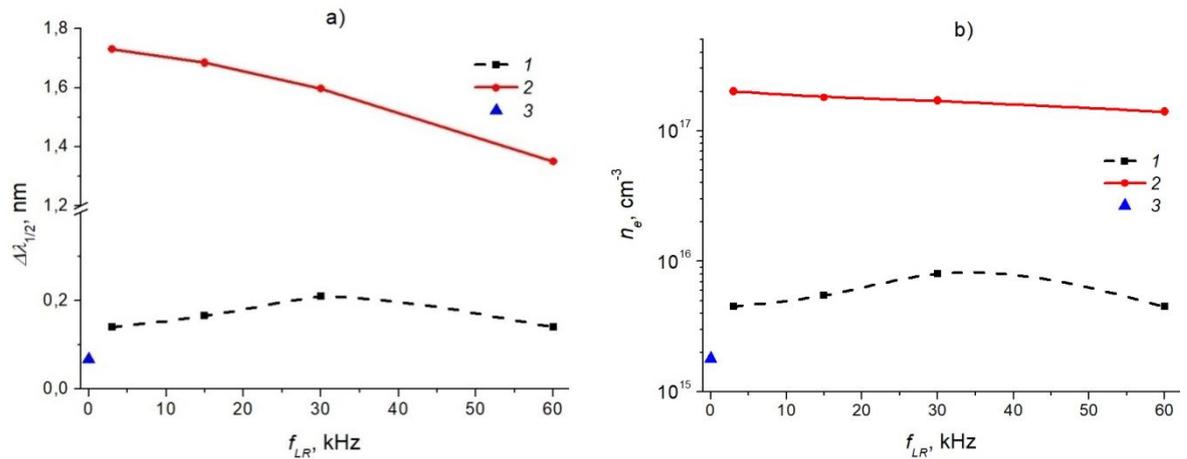

Fig. 4. Half-width of the $H_\alpha$ line (a) and the electron density (b), measured from the half-widths of the narrow – *1* and wide – *2* contours of the two-contour Lorentzian approximation of the $H_\alpha$ line shape in a hybrid plasma in the mixture of 89Ar:10H$_2$:1CH$_4$ and in a microwave plasma – *3* (Lorentzian shape).

Thus, it was shown that the electron density in the microwave discharge plasma in the 89Ar:10H$_2$:1CH$_4$ mixture at atmospheric pressure is ~$1.5 \cdot 10^{15}$ cm$^{-3}$. Introducing CO$_2$ laser radiation into the reactor chamber, in addition to the microwave radiation, leads to the formation of the hybrid plasma with a higher electron density. The electron density in the hybrid plasma, measured by the Stark broadening of the H$_\alpha$ lines, lies in the range from $4 \cdot 10^{15}$ cm$^{-3}$ to $2 \cdot 10^{17}$ cm$^{-3}$ and is determined by the spatiotemporal characteristics of the plasma, which depend on the laser pulse parameters and the localization of the CO$_2$ laser radiation focus relative to the microwave plasma bunch.

**Financial support**

The work was supported by the Russian Science Foundation grant No. 25-29-00816.

**Conflict of interest**

The authors declare that they have no conflict of interest.